\RequirePackage{amsmath}
\RequirePackage{amssymb}
\documentclass[runningheads]{llncs}

\usepackage{graphicx}
\usepackage{xspace}

\usepackage{cite}

\usepackage{xcolor}
\definecolor{comment}{rgb}{0.2, 0.6, 0.4}
\definecolor{string}{rgb}{0, 0, 0.5}
\definecolor{keyword}{rgb}{0.6, 0.2, 0.4}
\definecolor{background}{gray}{0.97}
\definecolor{lightgray}{gray}{.8}
\definecolor{lightblue}{rgb}{0.8,0.85,1}
\usepackage[T1]{fontenc}
\usepackage{pdfpages}

\usepackage{hyperref}
\usepackage{listings}
\usepackage{float}
\newfloat{lstfloat}{htb}{lop}
\floatname{lstfloat}{Listing}
 
\usepackage{lineno}

 \newcommand{\support}[1]{\mathit{sup}\!\left(#1\right)}
 \newcommand{\confidence}[1]{\mathit{conf}\!\left(#1\right)}
 \newcommand{\eat}[1]{}

 \newcommand{\bi}{\begin{itemize}}
 \newcommand{\ei}{\end{itemize}}

\newcommand{\xml}{\textsc{Xml}\xspace}

\newcommand{\declare}{\textsf{Declare}\xspace}

\newcommand{\prolog}{Prolog\xspace}
\newcommand{\datalog}{Datalog\xspace}
\newcommand{\swi}{\textsc{Swi}\xspace}

\newcommand{\python}{Python\xspace}

\newcommand{\red}[1]{#1\xspace}

\newcommand{\mbxl}[2]{\makebox[#1mm][l]{#2}}
\newcommand{\mbxr}[2]{\makebox[#1mm][r]{#2}}

\newcommand{\eg}{e.\,g.,\xspace}
\newcommand{\cf}{cf.\xspace}
\newcommand{\ie}{i.\,e.,\xspace}

\begin{document}

\title{Finding Maximal Non--Redundant \texorpdfstring{\\}{}
   Association Rules in Tennis Data}

\author{%
   Daniel Weidner\inst{1} \and
   Martin Atzmueller\inst{2} \and
   Dietmar Seipel\inst{1}}

\authorrunning{Weidner, Atzmueller and Seipel}

\institute{%
   University of W\"urzburg, Department of Computer Science, \\
   Am Hubland, 97074 W\"urzburg, Germany \\
   \email{\{daniel.weidner,dietmar.seipel\}@uni-wuerzburg.de} \\
   [.8ex] \and
   Tilburg University, Department of Cognitive Science \&
   Artificial Intelligence, \\
   Warandelaan 2, 5037 Ab, Tilburg, The Netherlands \\
   \email{m.atzmuller@uvt.nl}
}

\maketitle
\begin{abstract}
The concept of association rules is well--known in data mining.
But often redundancy and subsumption are not considered,
and standard approaches produce thousands or even millions
of resulting association rules.
Without further information or post--mining approaches, this huge
number of rules is typically useless for the domain specialist --
which is an instance of the infamous pattern explosion problem.

In this work, we present a new definition of redundancy and
subsumption based on the confidence and the support of the rules
and propose post--mining to prune a set of association rules.

In a case study, we apply our method to association rules
mined from spatio--temporal data.
The data represent the trajectories of the ball in tennis
matches -- more precisely, the points/times the tennis ball
hits the ground.
The goal is to analyze the strategies of the players and to try
to improve their performance by looking at the resulting
association rules.
The proposed approach is general, and can also be applied to other
spatio--temporal data with a similar structure.

\end{abstract}

\section{Introduction}
\label{sec:1}

The field of artificial intelligence (AI) can be divided
into symbolic and subsymbolic approaches,
\eg~\cite{smolensky1987connectionist,
mcmillan1992rule,goertzel2012perception,battaglia2018relational}.
\emph{Symbolic, knowledge-- or rule--based} AI models
central cognitive abilities of humans
like \emph{logic}, deduction and planning in computers;
mathematically exact operations can be defined.
\emph{Subsymbolic} or statistical AI tries to learn a model of a
process (\eg an optimal action of a robot or the classification
of sensor data), from the data.

Association rules declaratively and symbolically describe
logical relations with probabilities in the form of
if--then--rules, thus incorporating aspects from both symbolic and
statistical approaches. Then, using declarative specifications,
\eg using domain knowledge, specific (inductive) biases,
and post--mining approaches, the learning and mining can be
supported~\cite{atzmueller2007declarative,battaglia2018relational},
and post--mining on the set of association rules -- for improving
their interestingness and relevancy --
can be conveniently implemented.
In general, data mining aims to obtain a set of novel,
potentially useful and ultimately interesting patterns
from a given (large) data set~\cite{fayyad1996data}.
Here, one prominent method is
association rule mining. However, many standard approaches for
mining association rules -- like the well--known Apriori algorithm --
do not consider redundancy or subsumption.

In this paper, we tackle this problem, and demonstrate its
application in the spatio--temporal domain of tennis data.
Our contributions are summarized as follows:
We introduce a new definition of redundant and subsumed
association rules to prune the set of rules we obtain from
the Apriori algorithm.
Furthermore, we present a general post--mining approach for finding
maximal non--redundant association rules.
Based on the results of previous mining steps, unimportant
attributes are excluded in further steps.
%

\begin{quote}
\begin{figure} \center
\includegraphics[scale=0.35]{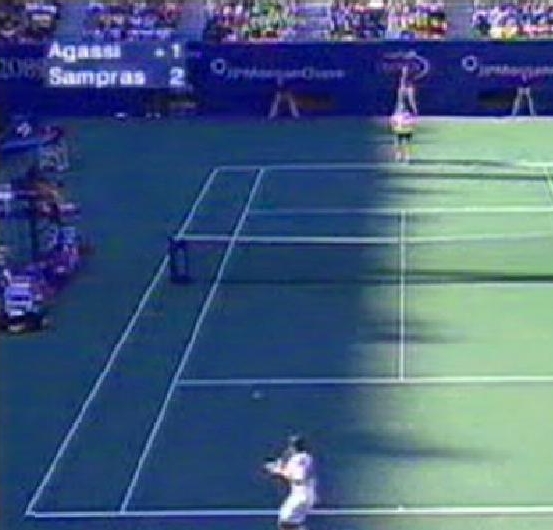}
\caption{Camera Shot of a Tennis Match.}
\label{agassisampras}
\end{figure}
\end{quote}

From an application perspective, analyzing real--world tennis data and pruning the result effectively can
lead to individual training methods for the observed players.
Furthermore, using data from a video in real--time,
a coach could change the player's strategy during
a tennis match.
Some pre--research had been made in the diploma/master
theses~\cite{wehner2003,baumgart2019}
and the technical report~\cite{seipelanalys};
by getting information directly from media--data like
video--sequences (a screenshot can be seen in Figure~\ref{agassisampras}),
essentially the complete (tennis) data mining process can now be
automated.

\eat{
From Figure~\ref{agassisampras} we create with~\cite{wehner2003} a
court like Figure~\ref{greencourt}.
\begin{quote}
\begin{figure} \center
\includegraphics[angle=90,scale=0.6]{court}
\caption{Created Tennis Court.}
\label{greencourt}
\end{figure}
\end{quote}
}

The rest of this paper is structured as follows:
Section~\ref{sec:2} discusses related work.
 After that, Section~\ref{sec:3} presents the background.
Next, we present the proposed data mining process including
the new definition of subsumption and maximality
in Section~\ref{sec:4}. The case study about tennis data shows
its usefulness in Section~\ref{sec:5}.
Finally, Section~\ref{sec:6} concludes with a summary
and lists directions for future work.

\section{Related Work}
\label{sec:2}

This section discusses related work on association rule mining,
condensed representations,
and finally post--mining approaches on sets of association rules.

\subsection{Association Rule Mining}

Association rule mining~\cite{agrawal1993mining,agrawal1994fast}
has been established as a prominent approach in data mining
and knowledge discovery in databases, \cf~\cite{hipp2000algorithms}
for a survey. Several efficient algorithms have been proposed,
including, \eg the Apriori~\cite{agrawal1994fast} and the
FP--Growth~\cite{HPY:00} algorithms.

In the research on association rules and finding the most relevant
ones, many approaches have been discussed.
In~\cite{kotsiantis2006association}, the authors present a method
to extract rules on user--defined templates or time constraints.
In~\cite{hilderman2013knowledge}, association rules
are ranked. This happens with different interestingness measures.
Different formats or properties of association rules
are discussed in~\cite{brin1997beyond}.
Finally, the pruning of redundant
rules is presented in~\cite{cristofor2002generating}.
Constraint--based data mining also tackles the problem of redundancy
in association rule mining. For example, the approach presented
in~\cite{bayardo2000constraint} presents an approach for
constraint--based rule mining in large, dense databases,
focusing on the interestingness of specializations of rules
relative to their parent generalization with specific thresholds.

In contrast to the approaches discussed above,
we employ a standard method for association rule mining
(\eg Apriori) which is customized using logic programming,
such that the mining step can be re--iterated in a declarative way.

\subsection{Condensed Representations and Post--Mining Approaches}

Condensed representations of association rules for
reducing redundancy mainly focus on closed itemsets,
\eg~\cite{zaki2000generating,BPTSL:00}.
Furthermore, also research in the domain of formal concept analysis
has resulted in several algorithms, \eg~\cite{PBTL:99,STBPL01};
also \cf~\cite{Boulicaut:05}
for a survey on condensed representations.
Furthermore, \cite{fournier2012mining} presents a mining approach
for finding the top--k non--redundant association rules
using an approximation algorithm.

Considering post--mining methods,
\cite{zhao2009post} discusses several techniques
for effective knowledge extraction from association rules,
while~\cite{marinica2008post,marinica2010knowledge,
mansingh2011using} apply ontologies to facilitate
the post--processing of a set of association rules,
also including interaction with a domain expert.
Logic--based post--mining approaches include a technique where
patterns are filtered using constraints formulated with
answer set programming (ASP)~\cite{gebser2016knowledge}.

Similar to the approaches described above,
we also apply post--mining but using declarative techniques.
We apply formalizations of subsumption and redundancy for
declaratively shaping the association rules.
However, specifically in contrast to the existing logic--based
approaches, the presented approach is not restricted to work
on the set of association rules directly,
but can further refine the mechanism of how to discover
association rules, by \eg refining the data representation,
the parameters of the mining process,
and its subsequent results at the same time in incremental fashion.

\section{Association Rules and the Apriori Algorithm}
\label{sec:3}

In this section, we provide an overview on the relevant background
on association rules, before briefly summarizing
the Apriori algorithm.

\subsection{Association Rules}
\label{sec:31}

We consider a set of transactions, where each transaction is a
set of items, called an itemset.
An \textit{association rule}
$r = L \Rightarrow R$ is a classification rule, where the
antecedent $L$ and the conclusion $R$ are itemsets;
\red{without loss of generality, we assume
   \( L \cap R = \emptyset. \)
}
The \textit{support} of an association rule is the number
of the transactions containing both sides \red{divided 
by the number of all transactions}.
The \textit{confidence} of an association rule expresses
the likelihood that $R$ occurs in a transaction, if $L$
occurs in the transaction. It is defined as the percentage
of transactions containing $L \cup R$ among the transactions
containing $L$.
We write $\support{r}$ and $\confidence{r}$
for the support and confidence, respectively.
\red{Note, that support and confidence do not depend on each other,
and both definitions are necessary for association rule mining.
There can be rules with a large support but a
small confidence, and vise versa.}

The main goal of association rules mining is
to find rules having a minimum confidence and support.
These rules may be obvious for an expert, but we will show
in a case study for tennis data that they can reveal new,
unknown relations.

\subsection{The Apriori Algorithm}\label{sec:32}

The Apriori algorithm is a standard method for finding
association rules according to specified minimal support and
minimal confidence thresholds.
This algorithm incrementally searches for
frequent itemsets, utilizing the minimal support threshold.
It starts with itemsets of size 1, and iteratively refines
the itemsets, enlarging the itemsets.
The algorithm stops, if there is no frequent itemset of
a certain size. From all frequent $k$--itemsets, all possible
($k+1$)--itemsets are considered if they are also frequent,
\ie they exceed the minimum support.

In the context of this paper, we apply a specific implementation
of the Apriori algorithm, provided by the well--known data mining tool
Weka~\cite{WEKA}. In Section~\ref{sec:52},
we will see, that Weka can be used with a parameter which
stands for a required number of rules. So in the Apriori
algorithm implemented in Weka, the support is reduced until
this number of rules is reached.
Depending on the size and characteristics of the data set,
this can potentially be a very large number of rules,
that also exceeds a given minimal confidence.
Depending on the size of the table and the minimal confidence,
this number of rules can be unusably high,
so that users lose track of those rules.
At this point, the process has to be run again with other parameters
or some expert needs to filter important rules, which ends in
looking for a needle in a haystack.
Since both of these methods are very time--consuming and expensive,
we propose a new approach by pruning rules effectively.
For this, we employ the idea of redundancy and subsumption,
which we define in the next section in detail.

\section{Data Mining Process}
\label{sec:4}

The proposed declarative data mining process,
which will be discussed in this section,
is implemented within the software package
\declare~\cite{Declare} for knowledge--based intelligent systems
that is developed using \swi--\prolog~\cite{swiref}.
We introduce a (semi--) automatic data mining process
that consists of the steps outlined below.
By repeating these steps with different parameters
or transformations, we achieve different results in each iteration,
until some result is convincing enough
according to the assessment of a domain expert.
Thus, the evaluation of the results has to be supported by
a domain expert or some knowledge structure like a
knowledge graph.
In the latter case, the process can then also be potentially
automatized,
\red{which we plan to do in the future; so far
the process workflow is semi--automatic.}
The process workflow is presented in Figure~\ref{flowchart}.
In the following, we will present the steps of an iteration.
\begin{quote}
\begin{figure} \center
 \includegraphics[scale=0.82]{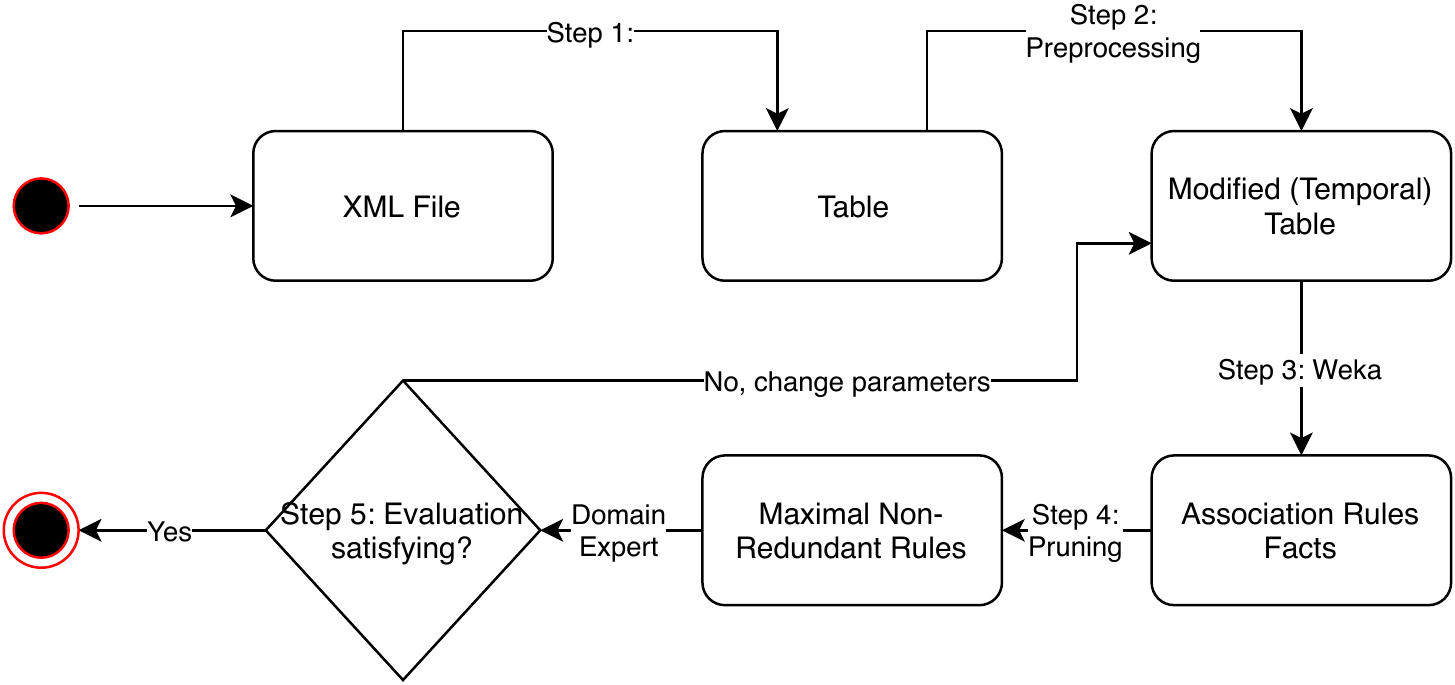}
 \caption{Flowchart of the Data Mining Process.}
\label{flowchart}
\end{figure}
\end{quote}

\subsection*{Step 1 and 2:
   From \xml to Modified (Spatio--Temporal) Table (Feature Selection and Extraction)}

Given an \xml file with data to be analyzed,
we want to create a structured table.
For this, some aspects should be discussed. First of all,
we have to think about which attributes of the \xml file
should be included in the table and the data mining process.
In our test data we skip some unimportant attributes
by projecting the table.
By now, the selected attributes are chosen first,
then the algorithm is run.
In the future we plan to make this interactive,
so the user is asked,
which attributes should be involved in one iteration.
As it can be seen in Section~\ref{sec:52}, we did not just
use the attributes of the \xml file, we also transform
attributes, which are exact coordinates, so the size of
the domain equals the size of the table, and the
attributes appear with a probability near to zero.
Another transformation is done to create a temporal data
scheme. Both transformations may be applicable to other
\xml files, but it is impossible to guide this automatically.
Nevertheless, if users are familiar with the used file, they
should think about such transformations, since attributes
with a probability near to zero will only reach a small support and
confidence, and a temporal data table allows other data
mining methods too, see~\cite{antunes2001temporal}.

\subsection*{Step 3: From a Table to Association Rules (via Weka)}

As discussed in Section~\ref{sec:32}, we use the
Apriori algorithm of Weka to get association rules.
This step includes two thresholds that can be modified in
each iteration: first the number of required rules can be
increased, and second the value of the minimum confidence
can be decreased. This may lead to a greater number of rules:
so the bigger the search space of rules, the higher is
the chance to find very interesting rules.

\subsection*{Step 4:
   Pruning Non--Maximal Redundant Association Rules}

After all found association rules are loaded into the system,
we want to get rid of redundant and non--maximal association rules.
For this, we need a new definition of redundancy,
where an association rule
   \( r_1 = L_1 \Rightarrow R_1 \)
is called \textit{redundant}, if there is another association rule
   \( r_2 = L_2 \Rightarrow R_2, \)
such that
\begin{quote}
   \( L_2 \subseteq L_1,\
      R_1 \subseteq R_2 \mbox{ and }
      \mathit{conf}\!\left(r_2 \right) = 1. \)
\end{quote}
\red{
Note that, if a rule $r_1$ is redundant,
then its confidence does not have to be $1$ in general;
e.g., for a redundant rule $r_1 = L_1 \Rightarrow R_1$
and a rule $r_2 = L_2 \Rightarrow R_2$ with
   \( \confidence{r_2} = 1, \)
such that $L_2 = L_1$ and $R_1 \subsetneq R_2$,
we have
   \( \emptyset = L_2 \cap R_2 = L_1 \cap R_2, \)
and we get
   \( 1 = \confidence{r_2} =
      \frac{\left| L_2 \cup R_2 \right|}{\left| L_2 \right|} =
      \frac{\left| L_1 \cup R_2 \right|}{\left| L_1 \right|} >
      \frac{\left| L_1 \cup R_1 \right|}{\left| L_1 \right|} =
      \confidence{r_1}. \)
}
Our definition of redundancy is different from related literature
like~\cite{kotsiantis2006association,
fournier2012mining,fournier2013tns}.

We say that an association rule $r_1 = L_1 \Rightarrow R_1$
subsumes another association rule $r_2 = L_2 \Rightarrow R_2$,
in short $r_1 \trianglerighteq r_2$, if
\begin{quote}
   \( L_1 \subseteq L_2,\
      R_2 \subseteq R_1 \mbox{ and }
      \mathit{sup}\!\left( r_1 \right)
         \geq \mathit{sup}\!\left( r_2 \right),\
      \mathit{conf}\!\left( r_1 \right)
         \geq \mathit{conf}\!\left( r_2 \right). \)
\end{quote}
A rule $r$ is called \textit{maximal}, if it is not subsumed
by any other rule $r' \not= r$.
\red{Both definitions have first appeared in the lecture~\cite{adbseipel};
like support and confidence they do not depend on each other.
This means, a subsumed rule can be non--redundant,
and a redundant rule may not be subsumed by any other rule.}
After applying these definitions, we hope to finally obtain a
small number of maximal non--redundant rules.


In the following, we motivate these definitions.
Essentially, to find interesting association rules,
these definitions are necessary.
This is also shown by Table~\ref{min_conf}, where we
ran Weka with the same table as in Section~\ref{sec:5},
but searching for all possible association rules.
This means the minimum confidence is increased by $0.1$,
and we count the number of all association rules found by Weka.
For an example with tennis data, which will be presented
in detail in Section~\ref{sec:5}, we found about $22\, 000$
rules, even if we required a minimum support of $1$.
And with a standard minimum confidence of $0.50$ or $0.75$,
we reach $70\, 000$ or $40\, 000$ rules.
As said before, searching for the set containing the most interesting
rules of this large total rule-set is nearly impossible;
but it is reasonable to work with a small number
of maximal non--redundant rules to obtain unknown relations.
\begin{quote}
\begin{figure} \center
\includegraphics[scale=0.7]{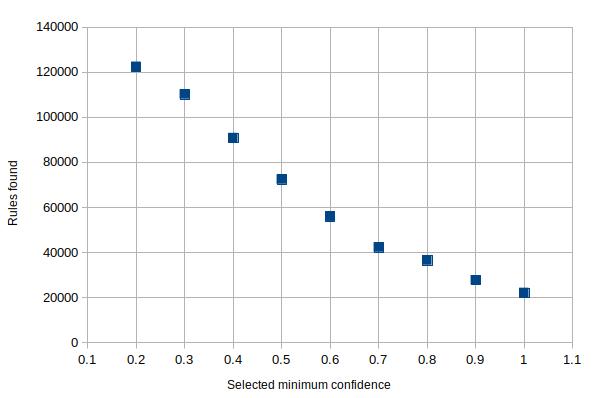}
\caption{Relation between Minimum Confidence
   and Number of Found Association Rules in the Tested Tennis Data Set.}
\label{min_conf}
\end{figure}
\end{quote}

\subsection*{Step 5: Iterating the Process with New Parameters}

After one iteration, some evaluation of the result
has to be done. Depending on the quality of the
evaluation, the system should repeat the process
with different parameters, starting with those
for Weka. This is because these thresholds need
no information about the initial data. It can
be increased and decreased, respectively, to create
more association rules.
If all these iterations fail the evaluation,
then some modification of the starting table and
attributes should be done.
For this, the user needs information or knowledge about the data
to guide the data mining process,
\ie in-- and excluding the right attributes and changing
the right parameters correctly.

\section{Case Study: Analysis of Tennis Data}
\label{sec:5}

A system for the management and analysis of tennis data
had been started in \swi--\prolog in the diploma
thesis~\cite{wehner2003},
where an \xml representation had been developed
(see Listing~\ref{xmltennis} below),
and some simple analysis had been done.
Later, this analysis had been extended with a functionality
to query the data~\cite{seipelanalys} using the
\prolog--based \xml processing utilities of \cite{seipel2002processing}.
\prolog is very useful here, since knowledge bases with
semi--structured, symbolic data, such as relational, deductive,
\xml or semantic web data can be handled nicely with
\prolog~\cite{Bra:11,clocksin2003programming}.

In the following subsections,
we are \emph{refining} the proposed \emph{data mining process}
for deriving suitable association rules.
%
First, the \xml file with the tennis data is transformed into
a relational table.
Columns are created from the attributes of the file;
attributes can be omitted,
if they should not be involved in the mining process.
This initial table is transformed into a modified,
temporal table;
some of these modifications are not universally applicable,
but key ideas may be portable to other types of data.
First, we create a \emph{tessellation} for the tennis court,
since exact coordinates will be repeated with a probability
near to zero.
Second, we duplicate a part of the table in order to model
the data in a special way, such that traditional data mining
is lifted to \emph{temporal} data mining.

Then, the association rules are computed using the
Apriori algorithm for frequent itemsets of the tool Weka.
After wrapping the Weka output text file of the rules
into \prolog facts and consulting them,
the \emph{maximal non--redundant rules} are filtered
as the desired association rules.


\subsection{Preparing the Data: Creating and Duplicating the Table}
\label{sec:51}

We start with a given \xml file in the format of~\cite{wehner2003};
an example is given in Listing~\ref{xmltennis}.
Here the main information is saved in the (sub--) elements
\texttt{set}, \texttt{game}, \texttt{point} and \texttt{hit}.
These attributes will mostly form the columns of our table.
For a different file, a similar approach is conceivable.

\begin{quote} \samepage
\begin{lstlisting}[
   label = xmltennis,
   caption = {\xml File of a Tennis Match.},
   language = xml,
   keywords = {xml, version, encoding} ]
<?xml version='1.0' encoding='ISO-8859-1' ?>
<match>
  <player id="A" name="Sampras"/>
  <player id="B" name="Agassi"/>
  <result>
    <score set="1" player_A="6" player_B="3"/> ...
  </result>
  <match_facts>
    <tournament>US Open 2002</tournament> ...
  </match_facts>
  <set id="1" score_A="5" score_B="3">
    <game id="1" service="A" score_A="0" score_B="0">
      <point id="1" top="B" service="A" score_A="0"
         score_B="0" winner="A" error="0">
         <hit id="1" hand="forehand" type="ground"
            time="00:00:42" x="0.17" y="-12.07"/>
         <hit id="2" hand="backhand" type="ground" 
            time="00:00:44" x="-0.49" y="5.89"/>
         <hit id="3" hand="backhand" type="ground"
            time="00:00:46" x="-3.92" y="-3.42"/>
         <hit id="4" hand="forehand" type="ground"
            time="00:00:48" x="3.56" y="2.06"/>
      </point> ...
    </game> ...
  </set> ...
</match>
\end{lstlisting}
\end{quote}
This file is transformed into a relational table,
where all information is saved, see Figure~\ref{xmltotable}.
\begin{quote}
\begin{figure}[h] \center
\includegraphics[scale=0.34]{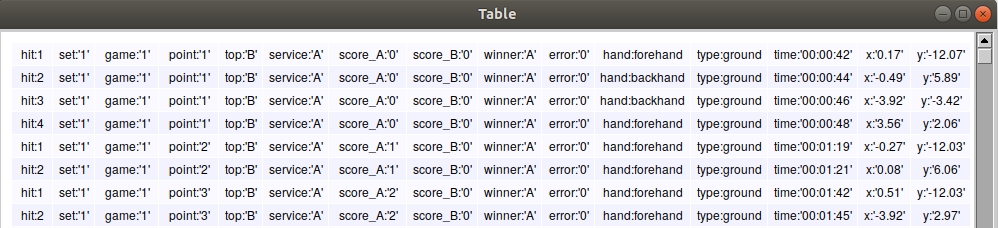}
\caption{Part of the Initial Relational Table for the Tennis Data.}
\label{xmltotable}
\end{figure}
\end{quote}
At this point, we do some small modification,
which is exclusive for the tennis data.
If we save the exact coordinates where the ball hits the ground,
then the same spot will be repeated
with a probability near to zero.
So we create a \emph{tessellation} for the court in $N \times M$
regions and instead of the exact \texttt{x}-- and \texttt{y}--coordinates
we save which intervals are reached, for example the red
box 1 inside the court. Note that also the outside of the court
forms intervals with numbers $1$ and $N+2$ or $M + 2$,
respectively, (\eg the blue box number 2),
see Figure~\ref{tessellation}.
\begin{quote}
\begin{figure}[h] \center
\includegraphics[angle=90,scale=0.335]{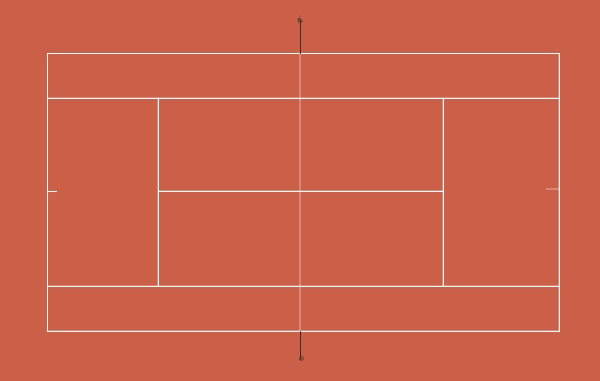}
\hfil
\includegraphics[scale=0.125]{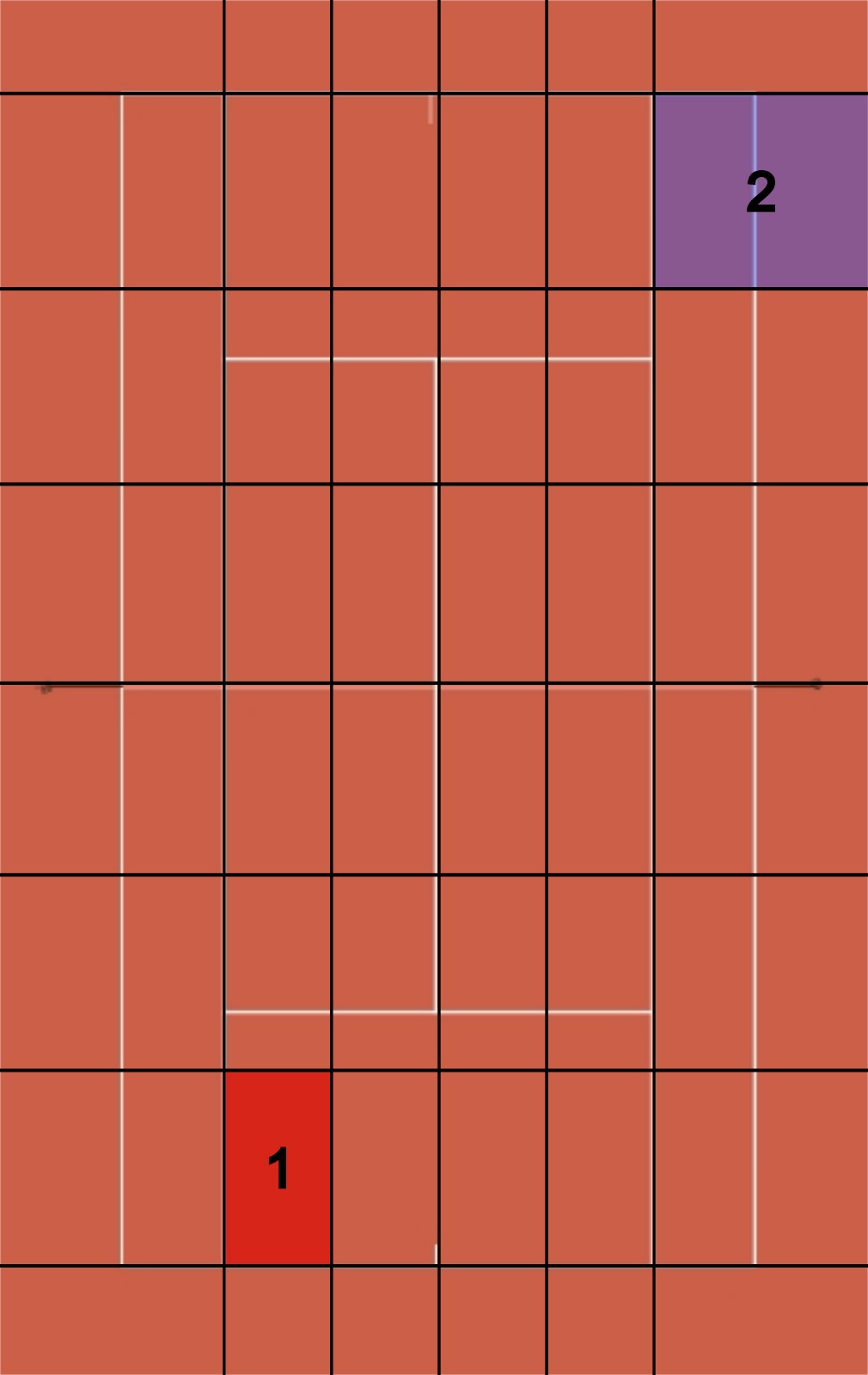}
\caption{Tessellation of a Tennis Court.}
\label{tessellation}
\end{figure}
\end{quote}
Next, we create the temporal table, by duplicating
part of the table of Figure~\ref{xmltotable}.
Here, we save the information for a hit and the next hit,
if there is one for this point.
This means, that if a point has a hit order 1-2-3,
we have the rows of hit pairs
$ \left( 1 , 2 \right) $ and $ \left( 2 , 3 \right) $.
In the resulting rows, all attributes are saved.
But, since not all information is useful, we omit some
which might not be interesting for the data mining process;
from Listing~\ref{xmltennis}, we skip \texttt{set},
\texttt{game}, \texttt{top}, \texttt{service}, \texttt{score\_A},
\texttt{score\_B};
also attributes for the temporal data, namely
\texttt{type\_\{1,2\}}, \texttt{time\_\{1,2\}},
\texttt{x\_\{1,2\}}, \texttt{y\_\{1,2\}}, are skipped.
Thus, we obtain the table of Figure~\ref{tabelletennis}.
Here the red boxes show the temporal data.
Note that the first attribute \texttt{hit} stands for hit pairs;
for example in the fourth point with the hits 1-2-3-4-5,
we get the four hit pairs $ \left( 1 , 2 \right) $,
$ \left( 2 , 3 \right) $,  $ \left( 3 , 4 \right) $ and
$ \left( 4 , 5 \right) $ (see the blue boxes).
\begin{quote}
\begin{figure} \center
\includegraphics[scale=1.8]{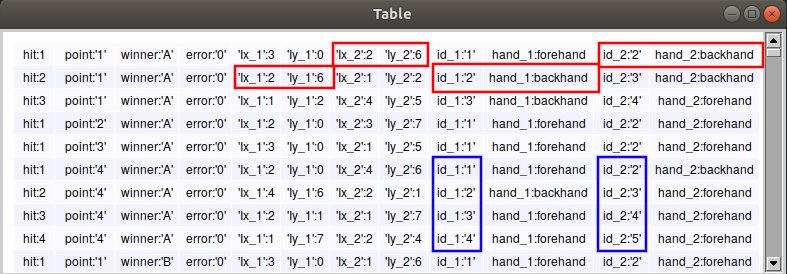}
\caption{Part of the Temporal Tennis Data.}
\label{tabelletennis}
\end{figure}
\end{quote}
This transformation could be modified,
such that we save hit triples. Also different numbers of
\texttt{x}-- and \texttt{y}--
intervals lead to different association rules.
The modification of the table may be supported by a domain expert.
In future work, we plan to involve some knowledge at this point.

\subsection{From Table to Association Rules}
\label{sec:52}

The input file \texttt{weka\_input.arff} of the data mining
tool Weka describes the attributes,
their domain and the rows of the table.
The other additional parameters \texttt{-N 4000}
and \texttt{-C 0.5} in the Weka call given
in Listing~\ref{wekacall} stand for
the required number of association rules (here 4000)
and the minimum confidence (here $0.5$), respectively;
their default values would be 10 and $0.9$.
Then we get an output text file \texttt{weka\_output}.
\begin{lstlisting}[
   label = wekacall,
   caption = {Bash Command to Call Weka.}]
java -cp ./weka.jar weka.associations.Apriori -t
   data/weka_input.arff -N 4000 -C 0.5 > data/weka_output
\end{lstlisting}
We call this bash command inside \prolog via the built--in
predicate~\texttt{unix/1} to create
the output, that looks like Listing~\ref{wekaoutput}.
Weka sorts the rules by the confidence and stops once
the number of required rules is reached.
\begin{quote}
\begin{lstlisting}[
   label = wekaoutput,
   caption = {Fragment of the Weka Output File.}]
Apriori
=======
Minimum support: 0.15 (409 instances)
Minimum metric <confidence>: 0.5
Number of cycles performed: 17
Generated sets of large itemsets:
Size of set of large itemsets L(1): 40 ...
Size of set of large itemsets L(9): 1

Best rules found:
1. service=B hand_1=forehand 1077 ==>
      type_2=ground 1077 conf:(1)
2. service=B hand_1=forehand type_1=ground 1072 ==>
      type_2=ground 1072 conf:(1)  ...
\end{lstlisting}
\end{quote}

\subsection{From Association Rules to Facts}
\label{sec:53}

From the output file obtained by Weka, we create corresponding
\prolog facts for the found association rules.
Since an association rule has four characteristic attributes,
namely antecedent, consequent, support and the confidence,
we save them in addition to a unique identifier. In particular
we get facts of the form
\begin{quote}
   \( \texttt{rule} ( \texttt{Id}, \texttt{Ant},
   \texttt{Cons}, \texttt{Sup}, \texttt{Conf}), \)
\end{quote}
where  \texttt{Ant}  and \texttt{Cons} are lists that
represent the itemsets.
These facts are obtained using the \prolog--based parsing and
\xml processing utilities of \declare.
After loading them into the system,
the user can query them in \prolog.
Currently, we are also experimenting with other programming
languages like \python for working with strings and text files.
\eat{%
We first wrote a wrapper in \prolog and later also in \python,
since it is a powerful language at working with strings and
text files. After skipping all lines up to
"\texttt{Best rules found:}" we read every line
and split them into the components from above.
This is done with \python commands split and find. After this we
write the rule into a new file \texttt{facts.pl}.
With consulting the created file 'facts.pl', we load all
rule facts into the system. Now the user can query those rules
like other \prolog facts.}

\subsection{From Facts to Maximal Non--Redundant Rules}
\label{sec:54}

From the facts for the association rules, we compute the
redundancy and subsumption in \declare.
Listing~\ref{declaredefintions} defines in \prolog
when the rule with the identifier \texttt{Id\_1} is redundant
because of the rule \texttt{Id\_2},
and when the rule with \texttt{Id\_1} is subsumed
by the rule \texttt{Id\_2}:
\begin{quote}
\begin{lstlisting}[
label = declaredefintions,
caption = {Definition of Redundant and Subsumed Rules in \declare.}]
redundant_rule(Id_1, Id_2) :-
   rule(Id1, Ant1, Cons1,_,_),
   rule(Id2, Ant2, Cons2,_,1),
   Id1 =\= Id2,
   subset(Ant2, Ant1), subset(Cons1, Cons2).

subsumed_rule(Id_1, Id_2) :-
   rule(Id_1, Ant1, Cons1, Sup1, Conf1),
   rule(Id_2, Ant2, Cons2, Sup2, Conf2),
   Id_1 =\= Id_2, Sup2 > Sup1, Conf2 > Conf1,
   subset(Ant2, Ant1), subset(Cons1, Cons2).
\end{lstlisting}
\end{quote}
For our example with the $4 \times 6$--tessellation,
these \prolog rules have derived 46~maximal non--redundant rules,
3 of which are given in Listing~\ref{result}.
\begin{quote}
\begin{lstlisting}[
label = result,
caption = {Example of Maximal Non--Redundant Rules.}]
41:  [id_2=2] => [hit=1,hand_1=forehand,id_1=1]
     Sup:0.31 Conf:1
42:  [id_1=1] => [hit=1,hand_1=forehand,id_2=2]
     Sup:0.31 Conf:1
416: [Iy_2=5] => [hand_1=forehand] Sup:0.18 Conf:0.88
\end{lstlisting}
\end{quote}
It is now an advanced task to find the rules,
which are useful for trainers or players.
Here, domain knowledge from experts has to be included.

\subsection{Experimental Results}
\label{sec:55}

In the following, some experiments and their results are discussed.
We compare the number of maximal non--redundant rules
with the maximal number of rules and the minimum confidence
in the Apriori algorithm of Weka.
\red{%
Table~\ref{experimental_results} shows that
the $4 \times 6$--tessellation leads to a small
(and so manageable) number of rules together with our
definitions of redundancy and subsumption.}
In most cases, less than 1\% of the required rules
are maximally non--redundant.
Only if we choose a minimum confidence of 0.5,
then we get up to 5\%.
Nevertheless, the total number of rules is useful,
and a domain expert has to check only a manageable number
of interesting rules.
Thus, the new definitions of subsumption and maximality
lead to convincing results for the tennis data.

\begin{table} \center
\begin{tabular}{c|c|c}
\rule[-1mm]{0mm}{4.5mm}%
Required Rules\ \ &\ Minimum Confidence\ \ &\
Maximal Non--Redundant Rules \\
\hline
\rule[-1mm]{0mm}{4.5mm}%
\mbxr{10}{5000} & \mbxl{10}{0.5} & \mbxr{10}{\textbf{46}} \\
\mbxr{10}{7500} & \mbxl{10}{0.5} & \mbxr{10}{\textbf{28}} \\
\mbxr{10}{10000} & \mbxl{10}{0.5} & \mbxr{10}{\textbf{156}} \\
\mbxr{10}{10000} & \mbxl{10}{1.0} & \mbxr{10}{\textbf{38}} \\
\mbxr{10}{12500} & \mbxl{10}{0.5} & \mbxr{10}{\textbf{442}} \\
\mbxr{10}{12500} & \mbxl{10}{1.0} & \mbxr{10}{\textbf{42}} \\
\mbxr{10}{15000} & \mbxl{10}{0.5} & \mbxr{10}{\textbf{773}} \\
\mbxr{10}{15000} & \mbxl{10}{0.75} & \mbxr{10}{\textbf{48}} \\
\end{tabular}
\vspace*{2mm}
\caption{Experimental Results.}
\label{experimental_results}
\end{table}

We have manually tested many different parameters for the
tessallations.
In the future, we would like to automatize the process of
searching for suitable parameters for convincing association rules,
which can be used as initial parameters in further experiments.
This will increase our experience about parameters leading
to good results.

\section{Conclusions and Future work}
\label{sec:6}

In this paper, we have introduced new definitions of redundant
and subsumed association rules, respectively.
With these definitions, we have discussed a data mining process.
In a case study, the definitions have proven useful
and led to results, which are far more convincing than the
results of the standard Apriori algorithm.

In the future, we are considering to use \datalog and
answer set programming for guiding an automatized data mining
process.
Again, the handling of the symbolic data (relations,
deduction or association rules) will be done in
logic programming.
In particular, we are planning to automatize parts of the
data mining workflow given in Figure~\ref{flowchart}
to decide on suitable parameters for the next iteration
based on an analysis of the previously derived association rules.
We will also consider other pattern mining approaches,
\eg subgroup discovery~\cite{wrobel1997algorithm,Atzmueller:15a}.

\bibliography{draft}
\bibliographystyle{plain}

\end{document}